\documentclass[aps,pre,twocolumn,groupedaddress,amsmath,amssymb]{revtex4}
\usepackage{graphicx,mathptm,bbm}
\usepackage{dcolumn}
\usepackage{bm}
\usepackage[colorlinks,linkcolor=red,citecolor=blue,urlcolor=blue]{hyperref}
\usepackage{amsthm,amsbsy}
\usepackage[usenames]{color}

\definecolor{DarkRed}{rgb}{0.7,0,0}

\begin{document}

\title{Coexistence of energy diffusion and local thermalization in nonequilibrium $XXZ$ spin chains with integrability breaking}

\author{J. J. Mendoza-Arenas$^1$, S. R. Clark$^{2,1}$ and D. Jaksch$^{1,2}$}
\affiliation{$^1$Clarendon Laboratory, University of Oxford, Parks Road, Oxford OX1 3PU, United Kingdom}
\affiliation{$^2$Centre for Quantum Technologies, National University of Singapore, 3 Science Drive 2, Singapore 117543}
\date{\today}

\begin{abstract}
In this work we analyze the simultaneous emergence of diffusive energy transport and local thermalization in a nonequilibrium one-dimensional quantum system, as a result of integrability breaking. Specifically, we discuss the local properties of the steady state induced by thermal boundary driving in a $XXZ$ spin chain with staggered magnetic field. By means of efficient large-scale matrix product simulations of the equation of motion of the system, we calculate its steady state in the long-time limit. We start by discussing the energy transport supported by the system, finding it to be ballistic in the integrable limit and diffusive when the staggered field is finite. Subsequently we examine the reduced density operators of neighboring sites and find that for large systems they are well approximated by local thermal states of the underlying Hamiltonian in the nonintegrable regime, even for weak staggered fields. In the integrable limit, on the other hand, this behavior is lost, and the identification of local temperatures is no longer possible. Our results agree with the intuitive connection between energy diffusion and thermalization.
\end{abstract}

\pacs{}

\maketitle

\section{Introduction} \label{intro}

In recent years the interest on the physics of nonequilibrium quantum systems has received a major impulse due to seminal developments in quantum simulation schemes~\cite{georgescu2014rmp,tommi_simulator}. In particular, ultracold atomic gases have emerged as some of the most attractive candidates to help unravel challenging questions on the physics of many-body interacting quantum systems~\cite{jaksch2005ann,bloch2008rmp,bloch2012nat}. Their high degree of controllability, isolation from the environment, and the existence of schemes for single-atom resolution~\cite{bakr2010sci,sherson2010nat}, make them ideal to simulate the physics of a vast variety of systems~\cite{bloch2008rmp,georgescu2014rmp}.

One of the most studied areas within the community of ultracold atomic gases corresponds to the dynamics of nonequilibrium interacting quantum systems~\cite{schneider2012nat,derrico2013njp,ronzheimer2013prl,vidmar2013prb}. Since the identification of the nature of transport supported even by testbed models of condensed matter systems is far from trivial, it is expected that their simulation in a highly controllable environment will help resolve several open questions. In particular, the relation between particle and energy transport through a quantum system and the integrability of its Hamiltonian, although intensively studied, is not fully understood. It has been shown that in integrable systems, the existence of nontrivial (local or quasilocal) conservation laws leads to ballistic conduction, as long as such laws have a finite overlap with the current operators~\cite{zotos1997prb,prosen2011strict,prosen2013quasi}. For nonintegrable models, in which nontrivial local conservation laws are absent, it is expected that a diffusion equation with finite conductivity is satisfied, i.e. that the transport is diffusive. Even though this is in fact the result found for several models~\cite{Heidrich-Meisner2003prb,jung2006prl,heidrich2007epjst,prosen2009matrix,benenti2009charge,langer2009real,karrasch2013prb,huang2013prb,karrasch2014prb,steinigeweg2014typicality,karrasch2014ladders}, ballistic transport in some nonintegrable systems has been reported~\cite{gobert2005real,langer2009real,karrasch2013prb}, or could not be ruled out~\cite{Heidrich-Meisner2003prb,heidrich2007epjst}.

The simulation of interacting systems in ultracold atomic gases could be determinant for establishing a definitive relation between integrability and transport~\cite{ronzheimer2013prl,vidmar2013prb}. A significant step towards this goal has been accomplished recently, due to the development of cold-atom configurations inducing particle transport through a mesoscopic channel connecting two reservoirs with population imbalance~\cite{brantut2012sci,stadler2012nat}. Moreover, by establishing different temperatures at the two reservoirs, thermoelectric effects have also been observed~\cite{brantut2013sci}. The use of these nonequilibrium configurations thus offers the possibility to study transport properties of quantum systems under widely differing conditions, with unprecedented control. 

A second problem whose research has been boosted by these experimental achievements with ultracold atomic gases is the relation between thermalization and integrability~\cite{polkovnikov2011rmp}. Specifically, it was suggested that for closed quantum systems taken to a nonequilibrium configuration, their local reduced density matrices do not relax to a thermal state if the Hamiltonian is integrable~\cite{weiss2006nat}, but tend towards a generalised Gibbs state incorporating the corresponding conservation laws~\cite{rigol2007prl}. On the other hand, several nonintegrable systems have been found to relax to a Gibbs state. A large amount of evidence indicates that this is achieved by means of a mechanism known as eigenstate thermalisation~\cite{rigol2008nat,rigol2009prl,steinigeweg2014prl2,steinigeweg2013pre,sorg2014pra}. However, these pictures are still under active debate~\cite{gogolin2011prl,rigol2012prl,sirker2014pra,posgay2014prl,wouters2014prl}. Moreover, thermalisation of open driven systems is much less well known, although bulk thermalisation in systems with nonintegrable Hamiltonian was found to be induced by thermal driving~\cite{znidaric2010pre}. 

Considering the impact of integrability on the transport and thermalization properties of quantum systems, the question of whether these phenomena are directly related naturally arises. Indeed, it would be expected that a system featuring ballistic transport does not tent towards a thermal state, due to the absence of scattering mechanisms which could equilibrate different parts of the system. It is also tempting to associate the relaxation towards a thermal state with diffusive transport, where dissipative mechanisms due to inelastic scattering take place. Even though this relation between transport and thermalization appears intuitive, it has yet to be explored. For example, just recently a connection between the relaxation towards a generalized Gibbs state and ballistic particle transport has been determined in a closed quantum system~\cite{mierzejewski2014prl}. To establish rigorously whether a general connection between the two types of phenomena actually exists, the coincidence of particular transport and thermalization regimes has to be shown first. In the present work we investigate the latter problem in a thermally-driven one-dimensional quantum system, extending the concept of local thermal states~\cite{hartmann2004prl,hartmann2004pre,hartmann2006cp,garcia_saez2009pra,ferraro2012epl,kliesch2014prx} to nonequilibirum configurations. As the main result of our work, we show the coexistence of diffusive energy transport and local thermalization in large nonintegrable systems. In the integrable regime, where ballistic energy transport emerges, local thermalization does not occur. 

The paper is organized as follows. In Section~\ref{model_section} we describe the model to be studied, corresponding to a spin chain thermally driven at its boundaries so an energy current is induced. In Section~\ref{direct_section} we discuss the properties of the energy transport resulting from a temperature imbalance across the spin chain, illustrating the transition between ballistic and diffusive regimes due to integrability breaking. Then we study in Section~\ref{local_thermal_section} the description of the thermally-driven system by means of local thermal states, and its relation to integrability. Our conclusions are presented in Section~\ref{conclu}.

\section{Model of boundary-driven system} \label{model_section}

\subsection{Spin chain model and boundary driving}

\begin{figure}
\begin{center}
\includegraphics[scale=1.08]{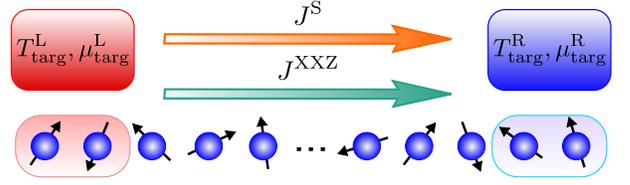}
\caption{\label{system_two_site_reservoir} (Color online) Scheme of the nonequilibrium system studied. At the left (L) and right (R) boundaries of a spin chain, thermal reservoirs of temperatures $T_{\text{targ}}^{\text{L,R}}$ and chemical potentials $\mu_{\text{targ}}^{\text{L,R}}$ induce local grand-canonical states on two neighboring spins. This leads to energy currents $J^{\text{XXZ}}$ (Eq.~\eqref{xxz_current_first}) and/or spin currents $J^{\text{S}}$ (Eq.~\eqref{spin_current}) through the chain. In the scheme, they flow from the left (red) to the right (blue) reservoir, assuming $T_{\text{targ}}^{\text{L}}>T_{\text{targ}}^{\text{R}}$ and/or $\mu_{\text{targ}}^{\text{L}}>\mu_{\text{targ}}^{\text{R}}$.}
\end{center}
\end{figure}

We start by describing the model to be considered in the present work, depicted in Fig.~\ref{system_two_site_reservoir}. The configuration consists of two thermal reservoirs of different temperature and/or chemical potential, located at the two edges of a one-dimensional spin chain. Due to the imposed imbalance, the chain is driven to a nonequilibrium steady state (NESS) supporting energy and/or spin currents. This setup is strongly motivated by the recent development of similar configurations in cold atomic systems~\cite{brantut2012sci,stadler2012nat,brantut2013sci}.

We describe the chain by the spin$-\frac{1}{2}$ $XXZ$ Hamiltonian, which corresponds to an archetypical model to analyze transport and thermalization properties of low-dimensional quantum systems~\cite{zotos1997prb,Heidrich-Meisner2003prb,jung2006prl,heidrich2007epjst,prosen2009matrix,benenti2009charge,langer2009real,znidaric2010dephasing,prosen2011strict,canovi2011prb,prosen2013quasi,karrasch2013prb,huang2013prb,we,we2,steinigeweg2013pre,we3,karrasch2014prb,sirker2014pra,steinigeweg2014prl2,steinigeweg2014typicality}. To investigate the effect of integrability breaking, we apply a staggered magnetic field in $z$ direction to the lattice~\cite{benenti2009charge,huang2013prb,steinigeweg2014typicality}. Thus the Hamiltonian is given by
\begin{equation} \label{hami}
H=\tau\sum_{j=1}^{N-1}(\sigma_j^x\sigma_{j+1}^x+\sigma_j^y\sigma_{j+1}^y+\Delta\sigma_j^z\sigma_{j+1}^z)+B\sum_{j=1}^N(-1)^j\sigma_j^z.
\end{equation}
Here $\hbar=1$, $\sigma_j^{\alpha}$ ($\alpha=x,y,z$) are the Pauli matrices at site $j$, $N$ is the number of sites, $\tau$ is the nearest-neighbor exchange coupling, $\Delta$ is the anisotropy parameter, which corresponds to the interaction strength between neighboring spin excitations, and $B$ is the amplitude of the staggered magnetic field. 

To study the nonequilibrium properties of the spin chain thermally driven at its boundaries, we follow the proposal of Refs.~\cite{prosen2009matrix,znidaric2010pre,znidaric2011jstat}, which allows for an efficient numerical simulation~\cite{zwolak2004mixed,cirac2004prl}. Specifically, we assume that the state of the system $\rho$ satisfies a Lindblad master equation
\begin{equation} \label{master_eq_general}
 \frac{d\rho}{dt}\equiv\mathcal{L}(\rho)=-i[H,\rho]+\mathcal{L}_{\text{L}}(\rho)+\mathcal{L}_{\text{R}}(\rho),
\end{equation} 
where the first term represents the coherent dynamics, and the dissipators $\mathcal{L}_k(\rho)$ correspond to the effect of the left ($k=\text{L}$) and right ($k=\text{R}$) reservoirs. Each superoperator $\mathcal{L}_k(\rho)$ is such that it induces a grand-canonical state of temperature $T$ and chemical potential $\mu$, namely
\begin{equation} \label{canonical_state_two}
\rho_2(T,\mu)=Z^{-1}e^{(-\varepsilon_{j,j+1}+\mu M_{j,j+1})/T},\quad Z=\text{Tr}(e^{(-\varepsilon_{j,j+1}+\mu M_{j,j+1})/T}),
\end{equation}
when acting on two spins $j, j+1$, with magnetization operator $M_{j,j+1}=\sigma_j^z+\sigma_{j+1}^z$, coupled by an $XXZ$ local Hamiltonian
\begin{align} \label{local_hami_thermal}
\begin{split}
\varepsilon_{j,j+1}&=\tau(\sigma_j^x\sigma_{j+1}^x+\sigma_j^y\sigma_{j+1}^y+\Delta \sigma_j^z\sigma_{j+1}^z)\\&+\frac{(-1)^jB}{2}[(1+\delta_{j,1})\sigma_j^z-(1+\delta_{j+1,N})\sigma_{j+1}^z].
\end{split}
\end{align} 
The reason for using these types of dissipators is that at least two sites are necessary to induce finite-temperature thermal states defined by the Hamiltonian couplings of interest (i.e. nearest-neighbor $XXZ$ interactions). Details of their implementation are given in Appendix~\ref{two_site_driving_app}. 

To drive the system to a nonequilibrium configuration, we apply these superoperators to its leftmost and rightmost pairs of spins, with \textit{target} temperatures $T_{\text{targ}}^{\text{L}}$ and $T_{\text{targ}}^{\text{R}}$ and chemical potentials $\mu_{\text{targ}}^{\text{L}}$ and $\mu_{\text{targ}}^{\text{R}}$ for the left (L) and right (R) boundaries~\footnote{The \textit{target} temperatures $T_{\text{targ}}$ and chemical potentials $\mu_{\text{targ}}$ are those that the reservoirs try to impose on the corresponding two boundary spins. For driving with no temperature imbalance, the values of the actual temperatures induced are, in most cases, approximately $2T_{\text{targ}}$, due to strong boundary effects~\cite{znidaric2011jstat}.}. The transport and thermalization properties of different sets of parameters are studied in the corresponding NESSs, obtained by simulating the long-time evolution of the system using the mixed-state time evolving block decimation algorithm~\cite{zwolak2004mixed,cirac2004prl}. This method allows us to reach system sizes much larger than those considered in previous studies of energy transport in interacting thermally-driven spin chains~\cite{mejia2007epjst,wichterich2007pre,michel2008prb,wu2011prb}. Our implementation is based on the open source Tensor Network Theory (TNT) library~\cite{tnt}. We note that our study is restricted to high temperatures ($T\gg\tau,\tau\Delta, B$), given that the calculations become considerably hard at low temperatures due to strong boundary effects and correlations~\cite{znidaric2011jstat}.

\subsection{Driving-induced NESSs and currents}

By selecting different target temperatures and chemical potentials, a large variety of effects can be studied. Namely, if $T_{\text{targ}}^{\text{L}}=T_{\text{targ}}^{\text{R}}$ and $\mu_{\text{targ}}^{\text{L}}=\mu_{\text{targ}}^{\text{R}}=0$, the steady state of the system does not show any net energy or magnetization flow, and thermalizes if the underlying Hamiltonian is nonintegrable~\cite{znidaric2010pre}. If a temperature imbalance is established, a NESS with an energy current is induced. The local energy current at site $i$ is given by the expectation value of the operator
\begin{equation} \label{xxz_current_first}
\begin{split}
J_i^{\rm XXZ}&=2\tau^2(\sigma_{i-1}^y\sigma_i^z\sigma_{i+1}^x-\sigma_{i-1}^x\sigma_i^z\sigma_{i+1}^y)\\&+\Delta\tau^2(\sigma_{i-1}^z\sigma_i^x\sigma_{i+1}^y-\sigma_{i-1}^y\sigma_i^x\sigma_{i+1}^z)\\&+\Delta\tau^2(\sigma_{i-1}^x\sigma_i^y\sigma_{i+1}^z-\sigma_{i-1}^z\sigma_i^y\sigma_{i+1}^x),
\end{split}
\end{equation}
as obtained from the continuity equation for the energy density in the bulk of the $XXZ$ spin chain~\cite{zotos1997prb}. If only a chemical potential imbalance is considered, with $\mu_{\text{targ}}^{\text{L}}=-\mu_{\text{targ}}^{\text{R}}$ and $T_{\text{targ}}^{\text{L}}=T_{\text{targ}}^{\text{R}}$, spin transport at zero average magnetization and finite~\cite{jesenko2011prb,znidaric2011jstat} or infinite~\cite{znidaric2010dephasing} temperatures can be simulated. In this case, the local spin current is given by the expectation value of the operator 
\begin{equation} \label{spin_current}
J_i^{\text{S}}=2\tau(\sigma_i^x\sigma_{i+1}^y-\sigma_i^y\sigma_{i+1}^x),
\end{equation}
obtained from the continuity equation of the local magnetization operator~\cite{zotos1997prb,benenti2009charge}. Furthermore, if there is a thermal or magnetization imbalance, and $\mu_{\text{targ}}^{\text{L}}\neq-\mu_{\text{targ}}^{\text{R}}$ so a finite magnetization is imposed to the system, magnetothermal effects arise, namely Seebeck and Peltier effects~\cite{louis2003prb,ajisaka2012prb,thingna2013epl}. This situation is briefly discussed in Appendix~\ref{magnetothermal_section} for the integrable limit, where we show that the nature of the magnetothermal response depends on the particular form in which it is induced.

Note that in the absence of bulk energy and magnetization dissipation, the energy and spin currents are homogeneous in the corresponding NESS~\cite{we2}. We thus denote them as $J^{\text{XXZ}}\equiv\langle J_j^{\text{XXZ}}\rangle/\tau^2$ and $J^{\text{S}}\equiv\langle J_j^{\text{S}}\rangle/\tau$ respectively.

\begin{figure}[t]
\begin{center}
\includegraphics[scale=0.7]{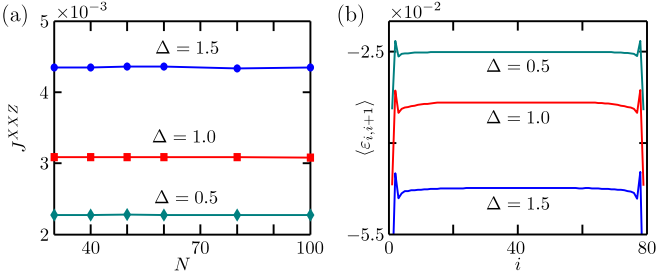}
\caption[Energy transport properties of integrable $XXZ$ spin chains.]{\label{ballistic_energy} (Color online) Energy transport properties of integrable $XXZ$ spin chains. The results correspond to $T_{\text{targ}}^{\text{L}}=\infty$, $T_{\text{targ}}^{\text{R}}=20$ and different interactions $\Delta$. (a) Energy current as a function of $N$. (b) Energy profiles for $N=80$. Note the strong boundary effects of the thermal driving.}
\end{center}
\end{figure}

\section{Direct energy transport and integrability} \label{direct_section}

We now consider the main question of our work, focusing on the impact of integrability breaking on the local properties of the NESS of thermally-driven systems. Thus during the rest of the paper we consider chains only driven by a temperature imbalance ($T^{\text{L}}_{\text{targ}}>T^{\text{R}}_{\text{targ}}$), with zero target chemical potentials. Also note that from here on, the numerical values of all the energies will be quoted in ratios of $\tau$, and for brevity the values of $B/\tau$, $T/\tau$, etc. will be referred to simply as $B$, $T$, etc. in figures and the main text.


We start our investigation by examining the nature of the direct energy transport through an $XXZ$ spin chain. We consider first the integrable case, with no staggered magnetic field ($B=0$). In Fig.~\ref{ballistic_energy}(a) we show for three interaction strengths $\Delta$ that the energy current through the system is independent of its size. In addition, we show in Fig.~\ref{ballistic_energy}(b) that the energy profiles are flat in the bulk. This indicates that the energy transport is ballistic for the different interaction regimes of the $XXZ$ model. This result thus provides strong evidence to support the picture of ballistic energy transport in integrable quantum systems, as discussed in previous works by means of different techniques~\cite{zotos1997prb,klumper2002jpa,orignac2003prb,louis2003prb,Heidrich-Meisner2003prb,heidrich2007epjst,langer2011prb,karrasch2013prb,karrasch2014prb}. Note that studies of energy transport in integrable systems with single-site thermal driving also suggested ballistic conduction, for $\Delta=0$~\cite{mejia2007epjst} and $\Delta=1$~\cite{manzano2012pre}, but for much smaller systems (of up to 12 sites).

\begin{figure}[t]
\begin{center}
\includegraphics[scale=0.7]{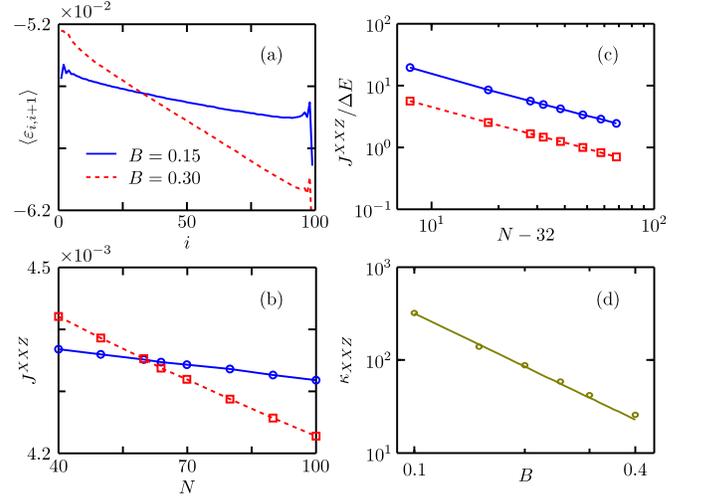}
\caption[Energy transport properties of the $XXZ$ model with staggered magnetic field.]{\label{energy_and_curr_stag} (Color online) Energy transport properties of the $XXZ$ model with staggered magnetic field. The simulations correspond to $\Delta=1.5$, $T_{\text{targ}}^{\text{L}}=\infty$ and $T_{\text{targ}}^{\text{R}} = 20$. (a) Examples of energy profiles for two staggered magnetic fields. (b) Corresponding energy currents as a function of $N$. The symbols represent the TNT results, and the lines are guides to the eye. (c) Scaling of the ratio $\langle J^{\text{XXZ}}\rangle/\Delta E$ with the size of the system. The symbols are the numerical data, and the lines represent the fits to equation~\eqref{diffusion_energy_staggered}. To perform the fit, we have discarded $n=15$ sites at each boundary of the chain for all values of $N$ considered. Larger values of $n$ do not modify the results, since the energy gradient is homogeneous in the region of the chain retained. For $B=0.15$, the fit gives $\kappa_{\text{XXZ}}=145(5)$ and $\alpha=0.98(1)$, and for $B=0.30$ it gives $\kappa_{\text{XXZ}}=40.6(5)$ and $\alpha=0.97(1)$. (d) Conductivity of the system as a function of $B$. The solid line corresponds to the fit $\kappa_{\text{XXZ}}=4.0(1.3)B^{-1.9(1)}$.}
\end{center}
\end{figure}

Next we consider the nonintegrable case with finite staggered magnetic field. First, we note that while the simulations for the case $B=0$ converged to the NESS quite fast, those of finite values of $B$ were found to be more demanding, with their convergence time scaling in a form $\sim B^{-1}$. For this reason we identified the amplitude $B=0.1$ as approximately the lowest one for which the NESS can be obtained with a reasonable computational effort. Thus we considered field amplitudes within the range $0.1\leq B\leq0.4$ for our study.

The most important features of the high-temperature energy transport of the nonintegrable system are shown in Fig.~\ref{energy_and_curr_stag}. We restrict the results to a single interaction strength, $\Delta=1.5$; a similar qualitative behavior was found for other $\Delta$ values. Specifically, as depicted in Fig.~\ref{energy_and_curr_stag}(a), the energy profiles are no longer flat, but acquire a ramp form that becomes steeper as $B$ increases. Also, as shown in Fig.~\ref{energy_and_curr_stag}(b), the energy current is no longer independent of the size of the system, but decreases with $N$. Thus the energy transport is no longer ballistic. Instead, as indicated in Fig.~\ref{energy_and_curr_stag}(c), it satisfies a diffusion equation in the bulk, namely
\begin{equation} \label{diffusion_energy_staggered}
\frac{J^{\text{XXZ}}}{\Delta E}=\frac{\kappa_{\text{XXZ}}}{(N-2n-2)^{\alpha}},\quad\Delta E=\langle\varepsilon_{N-n-1,N-n}\rangle-\langle\varepsilon_{n+1,n+2}\rangle
\end{equation}
with $\kappa_{\text{XXZ}}$ the energy conductivity, $n$ the number of sites discarded at each edge of the chain due to strong boundary effects~\cite{prosen2009matrix}, $\Delta E$ the energy difference between the leftmost and rightmost pairs of sites retained, and $\alpha\approx1$. In addition, as shown in Fig.~\ref{energy_and_curr_stag}(d), the energy conductivity diverges with the staggered magnetic field as $\kappa_{\text{XXZ}}\sim B^{-2}$ when $B\rightarrow0$, as expected from previous calculations~\cite{jung2006prl,huang2013prb}. So our results indicate that when the integrability of the Hamiltonian is broken, the energy transport becomes diffusive. This conclusion is consistent with recent calculations of current autocorrelation functions in systems with staggered magnetic fields~\cite{karrasch2013prb,huang2013prb,steinigeweg2014typicality}, and with previous studies in which the integrability is broken by means of other types of couplings~\footnote{Diffusive energy transport has been reported in systems with interchain couplings~\cite{Heidrich-Meisner2003prb,jung2006prl,karrasch2014prb} and next-nearest neighbor coupling leading to frustration~\cite{Heidrich-Meisner2003prb}. For dimerized systems, the nature of the energy transport is less clear, since different studies have obtained ballistic~\cite{karrasch2013prb} and diffusive~\cite{Heidrich-Meisner2003prb,heidrich2007epjst} regimes. Preliminary calculations with our nonequilibrium setup indicate diffusive energy conduction at high temperatures.}.  

We have therefore demonstrated, by using a transport scheme different to those considered in previous work, the existence of ballistic energy transport for an integrable Hamiltonian. On the other hand, the energy transport becomes diffusive when the integrability is broken, in this case by a staggered magnetic field. 

Now we examine the thermalization regimes in the same nonequilibrium configurations. We show the absence and emergence of thermalization on a local scale for sufficiently large chains with integrable and nonintegrable Hamiltonians, respectively, coinciding with ballistic and diffusive energy transport regimes.

\section{Local thermal states and integrability} \label{local_thermal_section}

An important problem regarding the nature of the NESS of a driven quantum system corresponds to whether, and under which conditions, it can be described by local equilibrium. If so, local temperatures and chemical potentials can be established, determining the simplest form in which a system can deviate from global equilibrium~\cite{prosen2010prl,benenti2014rmp}. In addition, considering the relation between relaxation to Gibbs-like states and nonlocal conservation laws in closed quantum systems~\cite{rigol2007prl,rigol2008nat,mierzejewski2014prl}, it becomes natural to ask whether Hamiltonian integrability is related to such a local equilibrium picture.

Here we study these questions by analyzing the concept of local thermalization in high-temperature thermally-driven systems. We find that the definition of local temperatures is possible in these configurations, depending on the integrability of the Hamiltonian. Namely, for large nonintegrable systems local thermalization arises, while it does not for integrable models.

\subsection{Correlation functions}

\begin{figure}[t]
\begin{center}
\includegraphics[scale=0.82]{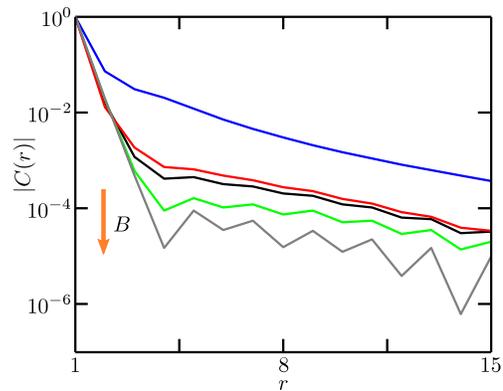}
\caption[Spin-spin correlations of the $XXZ$ model with staggered magnetic field.]{\label{correl} (Color online) Spin-spin correlations of the $XXZ$ model with staggered magnetic field, for $\Delta=1.5$, $N=100$, $T_{\text{targ}}^{\text{L}}=\infty$ and $T_{\text{targ}}^{\text{R}} = 20$. From top to bottom, the lines correspond to $B=0,0.15,0.20,0.30,0.35$. The field amplitude $B$ thus increases as indicated by the arrow.}
\end{center}
\end{figure}

A first point to evaluate regarding the possible existence of local thermal states in the NESS of the system is whether long-range correlations emerge. In Ref.~\cite{prosen2010prl} it was shown that when boundary driving induces spin transport at infinite temperature, long-range correlations emerge at interaction strengths $\Delta\gtrsim0.91$. At finite temperature long-range correlations were also found for $\Delta=1.5$. It was proposed that these results could demonstrate the  absence of well-defined local temperatures in nonequilibrium one-dimensional many-body systems. But interestingly, as discussed in the following Sections of our work, this turns out not to be the case in nonintegrable systems driven out of equilibrium by a thermal imbalance. Thus it is illustrative to observe first the behavior of spatial correlations across the system. In Fig.~\ref{correl} we plot the bulk-averaged correlation functions $C(r)=\langle C(j,r)\rangle_j$, with
\begin{equation}
C(j,r)=\langle\sigma_j^z\sigma_{j+r}^z\rangle-\langle\sigma_j^z\rangle\langle\sigma_{j+r}^z\rangle,
\end{equation}
as a function of the separation $r$ between spins. The notation $\langle.\rangle_j$ indicates spatial average of the correlations $C(j,r)$ with fixed $r$, excluding sites near the boundaries. The main observation from Fig.~\ref{correl} is that the correlations, which oscillate due to the staggered field, strongly decay with $B$, up to two orders of magnitude from $B=0$ to $B=0.4$. In addition, for $B>0$ the correlations are of $O(10^{-5})-O(10^{-6})$ for $r=15$, which indicates a much faster spatial decay than that of long-range correlations in Ref.~\cite{prosen2010prl}. This suggests that as the integrability-breaking parameter gets larger, a description of the system by means of local properties becomes more feasible.

\subsection{Determination of local thermal states} \label{determination_thermal}

\begin{figure}
\begin{center}
\includegraphics[scale=0.95]{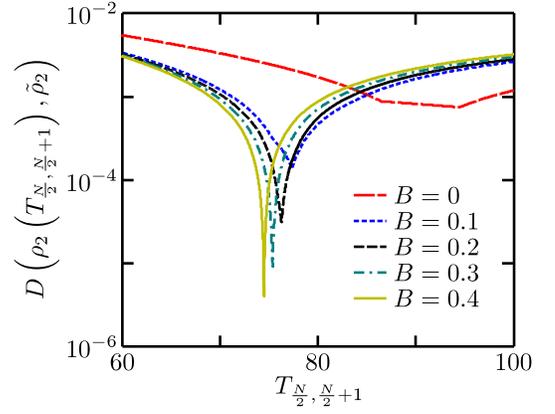}
\caption{\label{temp_nonequil_bvar} (Color online) Trace distance between the reduced density operator $\tilde{\rho}_2$ of the two central sites and two-site states~\eqref{grand_can_state_local} with temperature $T_{\frac{N}{2},\frac{N}{2}+1}$, for various staggered fields $B$. The calculations correspond to $N=100$, $T_{\text{targ}}^{\text{L}}\rightarrow\infty$, $T_{\text{targ}}^{\text{R}}=20$, $\Delta=1.5$, $\delta T=10^{-2}$, and the final iteration of the self-consistent procedure.}
\end{center}
\end{figure}

To determine whether the thermally driven system can be locally described by thermal states, we proceed as follows. First we calculate the reduced density operators of each pair of neighboring sites $(j,j+1)$ in the bulk of the driven system, which we denote as $\tilde{\rho}_2(j,j+1)$ \footnote{Reduced density operators are denoted by $\tilde{\rho}_n$, identified by the $\sim$ symbol. Thermal states are simply denoted as $\rho_n$. The sub-index $n$ refers to the number of sites of the density operators.}. Then we find the local two-site thermal state
\begin{align} \label{grand_can_state_local}
\begin{split}
&\rho_2(j,j+1)=Z_{j,j+1}^{-1}\exp\bigl(-(\varepsilon_{j,j+1}+\mu_j\sigma_j^z+\mu_{j+1}\sigma_{j+1}^z)/T_{j,j+1}\bigr),\\
&Z_{j,j+1}=\text{Tr}\Bigl[\exp\bigl(-(\varepsilon_{j,j+1}+\mu_j\sigma_j^z+\mu_{j+1}\sigma_{j+1}^z)/T_{j,j+1}\bigr)\Bigr],
\end{split}
\end{align}
with local temperature $T_{j,j+1}$ and chemical potentials $\mu_j$ and $\mu_{j+1}$, closest to $\tilde{\rho}_2(j,j+1)$. This state is identified by determining the free parameters $T_{j,j+1}$, $\mu_j$ and $\mu_{j+1}$ that minimize the trace distance~\cite{nielsen}
\begin{equation} \label{trace_dist}
D(\rho_2,\tilde{\rho}_2)=\frac{1}{2}\text{Tr}\left[\sqrt{(\rho_2-\tilde{\rho}_2)^2}\right].
\end{equation}
This calculation is performed self-consistently. First for each pair $(j,j+1)$ we fix the local chemical potentials to a particular value (see Eq.~\eqref{mf_chem_pots}), and sweep over a range of trial temperatures $T_{j,j+1}$ (with temperature step $\delta T$), as exemplified in Fig.~\ref{temp_nonequil_bvar} for the two central sites of the chain. The temperatures that minimize the trace distance are selected, and then are used to find new values of the local chemical potentials, following a similar minimization from a sweep over trial values. The process is repeated until convergence is obtained; see Appendix~\ref{self_consistent} for more details of this procedure. Finally, we compare expectation values of each $\tilde{\rho}_2$ with those of the closest thermal state found. If their difference is much smaller than the actual values of the expectation values (i.e. if the relative difference is small), $\tilde{\rho}_2$ corresponds to a local thermal state. 

A few points must be discussed before presenting our results. First, note that this method represents an improvement over procedures used in other works to find local temperatures~\cite{mejia2005epl,wu2011prb,znidaric2010pre}, which only relied on analyzing and comparing a few expectation values to determine thermalization. To understand why, consider two states $\rho$ and $\sigma$, and an observable $G$ with spectral decomposition $G=\sum_jg_j|j\rangle\langle j|$. If $p_j=\text{Tr}(\rho|j\rangle\langle j|)$ and $q_j=\text{Tr}(\sigma|j\rangle\langle j|)$ denote the probabilities of obtaining outcome $j$ in a measurement of $G$, the corresponding expectation values are
\begin{equation}
\langle G\rangle_{\rho}=\text{Tr}(\rho G)=\sum_jg_jp_j,\qquad\langle G\rangle_{\sigma}=\text{Tr}(\sigma G)=\sum_jg_jq_j.
\end{equation}  
Their difference is 
\begin{align} \label{trace_bounds_expect}
\begin{split}
|\langle G\rangle_{\rho}-\langle G\rangle_{\sigma}|&=\biggl|\sum_jg_j(p_j-q_j)\biggr|\leq|g^*|\sum_j|p_j-q_j|\\&\equiv2|g^*|D(p_j,q_j)\leq2|g^*|D(\rho,\sigma),
\end{split}
\end{align}
where $g^*$ is the eigenvalue of $G$ of maximal amplitude, $D(p_j,q_j)$ is the $L_1$ distance between the probability distributions $\{p_j\}$ and $\{q_j\}$, and where we have used that the trace distance $D(\rho,\sigma)$ upper-bounds $D(p_j,q_j)$~\cite{nielsen}. Thus the trace distance of two states upper-bounds the difference between the corresponding expectation values of \textit{any} observable (with finite eigenvalues). Its calculation then constitutes a well motivated measure of distance to determine the closest thermal state $\rho_2(j,j+1)$ to each $\tilde{\rho}_2(j,j+1)$. The sole value of the trace distance between both states, however, is not enough to determine whether $\tilde{\rho}_2(j,j+1)$ is actually thermal, since it does not give any indication of the relative difference between expectation values of the two states. This is why after finding the closest $\rho_2(j,j+1)$, a comparison of its expectation values to those of $\tilde{\rho}_2(j,j+1)$ is still required.

Second, it is important to discuss why we use the state of Eq.~\eqref{grand_can_state_local} to perform our study. An intuitive justification can be drawn from considerations on global thermal states at high temperature $T$. For such cases, where the total density operator is
\begin{equation} \label{high_T_global}
\rho_N=Z^{-1}\exp(-H/T)\approx\frac{1}{2^N}\left[I_N-\frac{1}{T}H+O\left(\frac{1}{T}\right)^2\right],
\end{equation} 
with $I_m$ the identity operator of $m$ sites, the reduced density operator of sites $j$ and $j+1$ is very well approximated by
\begin{align} \label{high_T_expansion_state}
\tilde{\rho}_{2}(j,j+1)&=\text{Tr}(\rho_N)_{(j,j+1)'}\approx\frac{1}{4}\left[I_2-\frac{1}{T}\varepsilon_{j,j+1}+O\left(\frac{1}{T}\right)^2\right]\notag\\&=Z^{-1}\exp(-\varepsilon_{j,j+1}/T),
\end{align}
with $Z$ the corresponding local partition function~\cite{garcia_saez2009pra}. 
The states of Eq.~\eqref{high_T_expansion_state}, however, do not account for the coupling of the pair of sites $(j,j+1)$ to the rest of the chain, even under some approximation. An initial improvement corresponds to assuming a mean-field (MF) coupling between the pair and the neighboring sites, namely
\begin{equation}
\sigma_{j-1}^{\alpha}\sigma_j^{\alpha}\approx\langle\sigma_{j-1}^{\alpha}\rangle\sigma_{j}^{\alpha}+\sigma_{j-1}^{\alpha}\langle\sigma_{j}^{\alpha}\rangle,
\end{equation}
for $\alpha=x,y,z$, and similarly for pair $(j+1,j+2)$. A reasoning similar to that of Eqs.~\eqref{high_T_global} and~\eqref{high_T_expansion_state} then leads to a state of the form in Eq.~\eqref{grand_can_state_local}, with
\begin{equation} \label{mf_chem_pots}
\mu_j=\tau\Delta\langle\sigma_{j-1}^z\rangle,\qquad\mu_{j+1}=\tau\Delta\langle\sigma_{j+2}^z\rangle,
\end{equation}
when considering that only $\langle\sigma_j^z\rangle\neq0$ when $B>0$. Thus the coupling of the two sites of interest to the rest of the chain motivates the inclusion of site-dependent chemical potentials on the local description of the NESS. To go beyond a mean-field approximation, these are taken as fitting parameters.

Even though we do not have a global thermal state but a system with temperature imbalance and energy transport, we consider states of Eq.~\eqref{grand_can_state_local} for our local analysis. This was further motivated by verifying numerically that the two-site reduced density operators of the $XXZ$ driven systems $\tilde{\rho}_2(j,j+1)$ have the form
\begin{equation} \label{expansion_two_site}
\tilde{\rho}_2=\frac{1}{4}\biggl(I_2+d_j\sigma_j^z+d_{j+1}\sigma_{j+1}^z+\sum_{\alpha=x,y,z}c_{j,j+1}^{\alpha}\sigma_j^{\alpha}\sigma_{j+1}^{\alpha}\biggr),
\end{equation}
with $d_j=\langle\sigma_j^z\rangle$ and $c_{j,j+1}^{\alpha}=\langle\sigma_j^{\alpha}\sigma_{j+1}^{\alpha}\rangle$. Since only the terms of Eq.~\eqref{expansion_two_site} are generated by the exponential of Eq.~\eqref{grand_can_state_local} \footnote{Its is readily verified from simple algebra that any $n$th power of $\varepsilon_{j,j+1}$ has the form $\varepsilon_{j,j+1}^n=C_1I_2+C_2\sigma_{j}^z+C_3\sigma_{j+1}^z+\sum_{\alpha}C_{\alpha,\alpha}\sigma_j^{\alpha}\sigma_{j+1}^{\alpha}$, with $C_{\alpha}$ and $C_{\alpha,\alpha}$ coefficients that depend on the Hamiltonian parameters. Thus the exponential of the Hamiltonian also has the same type of expansion.}, using it for the local description of a nonequilibrium setup stands as a very appealing and natural choice. Finally, note that since the operators describing the energy current correspond to three neighboring sites (see Eq.~\eqref{xxz_current_first}), they are not incorporated in our two-site description. However, they should be included in an analysis of the reduced density operators of more than two sites. Establishing a well motivated ansatz for the description of such reduced density operators remains an open question.

\subsection{Impact of integrability on local thermalization} \label{main_results_section}

\begin{figure}[t]
\begin{center}
\includegraphics[scale=0.95]{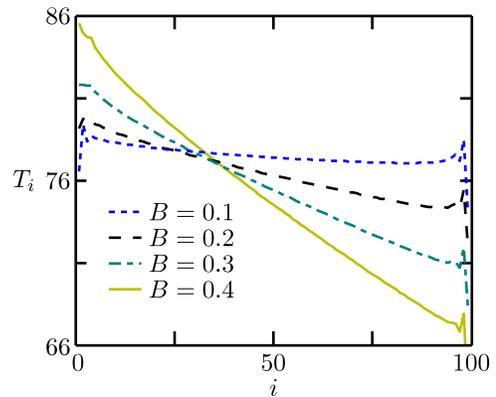}
\caption[Temperature profiles.]{\label{temp_profiles} (Color online) Temperature profiles across the system, for the parameters of Fig.~\ref{temp_nonequil_bvar}. The (flat) profile of $B=0$ is not shown since it corresponds to temperatures $\approx95$.}
\end{center}
\end{figure}

Now we discuss whether the two-site reduced density operators of thermally-driven systems of a fixed system size, namely $N=100$ spins, can be well approximated by the two-site thermal states of Eq.~\eqref{grand_can_state_local}. As described in Section~\ref{determination_thermal}, we start by identifying the local thermal state $\rho_2(j,j+1)$ closest to the reduced density operator $\tilde{\rho}_2(j,j+1)$ of each pair of neighboring spins of the driven chain. The determination of the local temperature is illustrated in Fig.~\ref{temp_nonequil_bvar} for the two central sites and various staggered magnetic fields $B$. Notably, the trace distance between the two types of states decreases and gets sharper as the staggered field $B$ increases. The resulting local temperatures for each value of $B$ are shown in Fig.~\ref{temp_profiles}. As expected, they describe well defined linear profiles. Note also that the obtained temperatures at the boundaries are significantly different to the target temperatures. This occurs due to the strong boundary effects of the two-site driving~\cite{znidaric2010pre,znidaric2011jstat}. In particular, the temperature at the left boundary is finite, while $T_{\text{targ}}^{\text{L}}\rightarrow\infty$. This results from the coupling of the two leftmost spins to the rest of the chain, which has finite local temperatures due to the finite value of $T_{\text{targ}}^{\text{R}}$. Additionally, observe that as $B$ increases the temperature profiles become steeper, an expected result since the system goes deeper into the diffusive regime. However, due to the different strength of boundary effects at each boundary (being stronger at low temperatures~\cite{znidaric2011jstat}), this steepening is asymmetric, resulting in the different temperature profiles crossing away from the center of the spin chain.

\begin{figure}[t]
\begin{center}
\includegraphics[scale=0.95]{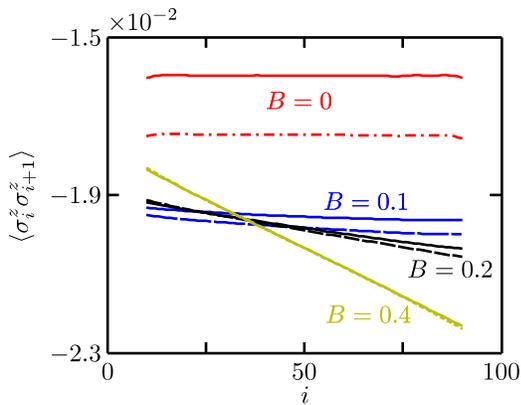}
\caption{\label{compar_xx_zz_nonequil} (Color online) Comparison between expectation values $\langle\sigma_i^{z}\sigma_{i+1}^{z}\rangle$ directly obtained from the numerical simulations of driven systems with temperature imbalance (dashed lines), and those of the chosen two-site thermal states (solid lines). Each indicated value of $B$ refers to the closest solid and dashed line. For clarity, the dashed lines correspond to: $B=0$ (dash-dot), $B=0.1$ (long-dashed), $B=0.2$ (medium-dashed) and $B=0.4$ (short-dashed). Also, the results for the ten leftmost and rightmost sites have not been plotted. The calculations correspond to the parameters of Fig.~\ref{temp_nonequil_bvar}.}
\end{center}
\end{figure}

Then we compare the corresponding $\langle\sigma_j^{\alpha}\sigma_{j+1}^{\alpha}\rangle$ expectation values of the two types of states. In Fig.~\ref{compar_xx_zz_nonequil} we present the comparison for $\alpha=z$; the results for $\alpha=x,y$ have the same features, so they are not shown. For $B=0$, the maximum difference of expectation values is $\approx9\%$. It is significantly diminished for $B=0.1$ ($\approx2\%$), and becomes very small for $B=0.4$ ($0.7\%$). These relative differences are consistent with the corresponding trace distance (see Eq.~\eqref{trace_bounds_expect}). Thus we conclude that away from the integrable limit, the NESS of the thermally-driven system of $N=100$ spins is locally well described by thermal states of the form in Eq.~\eqref{grand_can_state_local}. Close to and at integrability, this local description does not hold. 

\begin{figure}
\begin{center}
\includegraphics[scale=0.95]{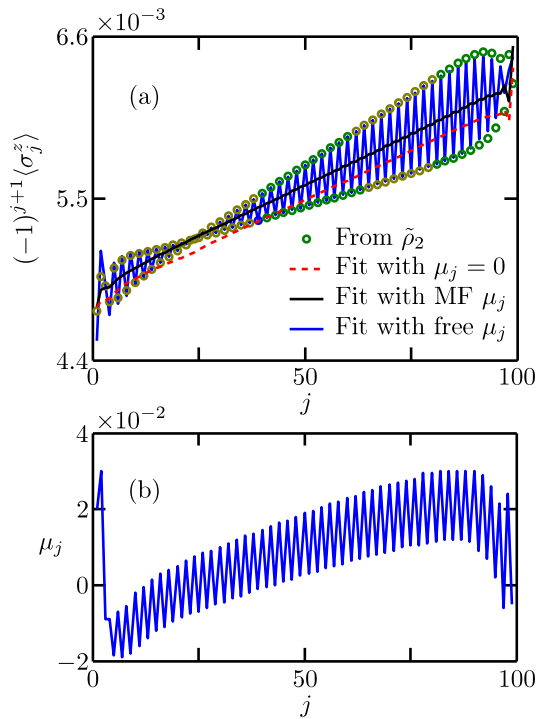}
\caption{\label{magnet_prof} (Color online) (a) Comparison between expectation values $\langle\sigma_j^{z}\rangle$ directly obtained from the numerical simulations of driven systems with temperature imbalance ($\circ$), and those of various two-site thermal states in Eq.~\eqref{grand_can_state_local}. The red dashed line refers to states with no local chemical potential. The black solid line corresponds to states with the mean-field chemical potentials in Eq.~\eqref{mf_chem_pots}. The solid blue (oscillating) line represents the results when using $\mu_j$ and $\mu_{j+1}$ as fitting parameters. (b) Values of $\mu_j$ minimizing the trace distance between the two-site reduced and thermal states of sites $(j,j+1)$. The calculations correspond to $B=0.4$, and the other parameters of Fig.~\ref{temp_nonequil_bvar}.}
\end{center}
\end{figure}

This conclusion is reinforced when looking at the magnetization profiles of the NESS. In Fig.~\ref{magnet_prof}(a) we show the staggered magnetization of the chain for $B=0.4$, along with the profiles obtained when fitting the local reduced density operators $\tilde{\rho}_2$ with different versions of Eq.~\eqref{grand_can_state_local}. Local thermal states with zero or mean-field chemical potentials reproduce the oscillatory form of the profile. However, the staggered magnetization has an additional staggering amplitude on top of it, which is not captured by any of these two limits. This additional residual staggering is reproduced well when taking $\mu_j$ and $\mu_{j+1}$ as free parameters, as seen in Fig.~\ref{magnet_prof}(a). The corresponding values of $\mu_j$ obtained for each pair of sites $(j,j+1)$ are shown in Fig.~\ref{magnet_prof}(b). In the bulk, these chemical potentials form an oscillating profile around a linearly increasing trend, resembling the increase of the magnetization profile.

On the other hand, for Hamiltonians close to integrability the magnetization values of the NESS are much lower, and cannot be reproduced even when $\mu_j$ and $\mu_{j+1}$ are free parameters. For example, for $B=0.1$, differences between the $\langle\sigma_j^z\rangle$ values of the NESS and the states in Eq.~\eqref{grand_can_state_local} minimizing the corresponding trace distance are of up to $20\%$ (not shown).

\subsection{Local states close to and at integrability}

\begin{figure}[t]
\begin{center}
\includegraphics[scale=0.95]{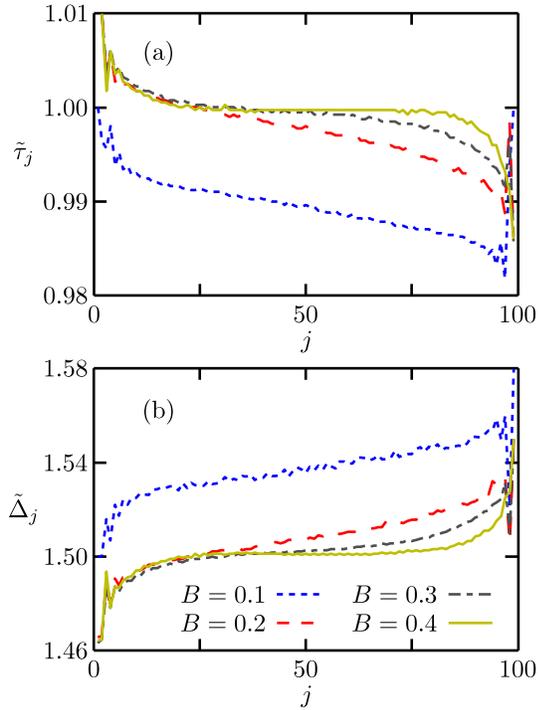}
\caption{\label{eff_couplings} (Color online) Effective site-dependent Hamiltonian couplings of local two-site states, for different staggered magnetic fields $B$, and interaction $\Delta=1.5$. (a) Effective hopping $\tilde{\tau}_j$. (b) Effective coupling in $z$ direction $\tilde{\Delta}_j$. The calculations correspond to the parameters of Fig.~\ref{temp_nonequil_bvar}. The results of $B=0$ are not shown since they are located in significantly different ranges of the $y$ axis, i.e. $\tilde{\tau}_j\approx0.93$ and $\tilde{\Delta}_j\approx1.77$ in the bulk.}
\end{center}
\end{figure}

We have noted in Section~\ref{determination_thermal} that the two-site reduced density operators $\tilde{\rho}_2$ of the thermally-driven $XXZ$ spin chains have the form specified in Eq.~\eqref{expansion_two_site}. Since this result holds independently of the value of the staggered magnetic field, it is natural to ask whether close to and at the integrable limit, the system can be described locally by states of the form of Eq.~\eqref{grand_can_state_local}, but with an effective local Hamiltonian 
\begin{align} \label{local_hami_effective}
\begin{split}
\tilde{\varepsilon}_{j,j+1}&=\tilde{\tau}_j(\sigma_j^x\sigma_{j+1}^x+\sigma_j^y\sigma_{j+1}^y+\tilde{\Delta}_j\sigma_j^z\sigma_{j+1}^z)\\&+\frac{(-1)^jB}{2}[(1+\delta_{j,1})\sigma_j^z-\sigma_{j+1}^z(1+\delta_{j+1,N})],
\end{split}
\end{align} 
where $\tilde{\tau}_j$ and $\tilde{\Delta}_j$ are free fit parameters corresponding to effective site-dependent hopping rates and interaction strengths, respectively (following the convention described in Section~\ref{direct_section}, the numerical values of $\tilde{\tau}_j/\tau$ are just denoted by $\tilde{\tau}_j$). We verified that this is in fact the case for several staggered magnetic field, including $B=0$. However, since the parameters $\tilde{\tau}_j$ and $\tilde{\Delta}_j$ obtained by this fitting deviate from the couplings of the parent Hamiltonian, there is no true local thermalization, and no local temperatures can be assigned to the system.

Specifically, we found for each two-site reduced density operator in the bulk of the system a state in Eq.~\eqref{grand_can_state_local} with effective local Hamiltonian of Eq.~\eqref{local_hami_effective}, so their trace distance is of $O(10^{-6})$. In Fig.~\ref{eff_couplings} we show the effective site-dependent couplings that minimize the corresponding trace distance for systems of size $N=100$, interaction $\Delta=1.5$ and various staggered fields. Deep in the nonintegrable regime ($B=0.4$), $\tilde{\tau}_j\approx1$ and $\tilde{\Delta}_j\approx1.5$ in the bulk. As the staggered field decreases, the effective parameters notably deviate from the values of the Hamiltonian couplings. Finally this deviation becomes very large in the integrable limit. In particular, $\tilde{\tau}_j\approx0.93$ and $\tilde{\Delta}_j\approx1.77$ for $B=0$ (not shown). These results indicate, in a complementary form to that of Section~\eqref{main_results_section}, that thermally-driven strongly nonintegrable systems are locally described by thermal states of the underlying Hamiltonian, while in the integrable limit such a description is not valid.

There are, however, three specific instances of integrability that require special attention, given that they satisfy the conditions described above to argue the existence of local thermalization. These correspond to the isotropic ($\Delta=1$), $XX$ ($\Delta=0$) and Ising ($\tau=0$, $\Delta\rightarrow\infty$) coupling limits. To explain what makes these cases special, and to show that their local two-site description by thermal states of the parent Hamiltonian is an artifact of their high symmetry, we take the first case as an example. Here, since all the directions are completely equivalent, the two-site reduced density operators $\tilde{\rho}_2(j,j+1)$ must have the form
\begin{equation} \label{rho2_tilde_Delta1}
\tilde{\rho}_2(j,j+1)=\frac{1}{4}\biggl(I_2+c_{j,j+1}\sum_{\alpha=x,y,z}\sigma_j^{\alpha}\sigma_{j+1}^{\alpha}\biggr),
\end{equation} 
with $c_{j,j+1}$ a local coefficient, equal for the three directions $\alpha$. Due to the symmetry of the $\Delta=1$ local Hamiltonian, given by
\begin{equation} \label{xxz_hami_only}
h_{j,j+1}=\tau(\sigma_j^x\sigma_{j+1}^x+\sigma_j^y\sigma_{j+1}^y+\sigma_j^z\sigma_{j+1}^z),
\end{equation}
it is easily shown that a two-site thermal state at temperature $T$ has the form \footnote{The key point to derive this result is that all the powers of the $\Delta=1$ local Hamiltonian have the form $h_{j,j+1}^n=\tau^n(a_n+b_nh_{j,j+1})$, with $a_n=3b_{n-1}$ and $b_n=(1-(-3)^n)/4$. Thus $\exp(-h_{j,j+1}/T)\propto h_{j,j+1}$.}
\begin{align} \label{rho2_Detla1}
\rho_2(j,j+1)=\frac{e^{-h_{j,j+1}/T}}{\text{Tr}(e^{-h_{j,j+1}/T})}=\frac{1}{4}(I_2+C(T)h_{j,j+1}),
\end{align}
with the coefficient
\begin{equation}
C(T)=\langle\sigma_j^{\alpha}\sigma_{j+1}^{\alpha}\rangle=\frac{e^{-\tau/T}-e^{3\tau/T}}{3e^{-\tau/T}+e^{3\tau/T}}.
\end{equation}
So by selecting the temperature that satisfies $\tau\,C(T)=c_{j,j+1}$, each two-site reduced density operator of the driven system is identified with a local thermal state with Hamiltonian $h_{j,j+1}$, in spite of the integrability. We have verified this result within our numerical simulations, finding trace distances between states $\rho_2(j,j+1)$ and the closest $\tilde{\rho}_2(j,j+1)$ of $O(10^{-7})$ in the bulk. Additionally, we have confirmed that $\tilde{\tau}_j=1$ and $\tilde{\Delta}_j=1$ when looking for the effective Hamiltonian couplings that minimize the trace distance. 

Similar arguments can be derived for the $XX$ and Ising limits. This is because $\langle\sigma_j^x\sigma_{j+1}^x\rangle=\langle\sigma_j^y\sigma_{j+1}^y\rangle\neq0$ and $\langle\sigma_j^z\sigma_{j+1}^z\rangle=0$ for the $XX$ chain, and only $\langle\sigma_j^z\sigma_{j+1}^z\rangle\neq0$ for the Ising model. As a result, both the two-site reduced density operators of the thermally-driven system and the two-site thermal states are proportional to the corresponding local Hamiltonian. By an appropriate selection of the local temperatures, the two types of local states coincide. 

There is, however, a key difference between the results for these particular integrable limits and those of nonintegrable Hamiltonians studied above, which justifies our conclusion that real thermalization emerges in the latter but not in the former. This is that our discussions for nonintegrable systems do extend to larger reduced density matrices. For instance, we have verified for $\Delta=1$ that when $T_{\text{targ}}^{\text{L}}=T_{\text{targ}}^{\text{R}}$, only the two-site reduced density operators correspond to thermal states with local Hamiltonian~\eqref{xxz_hami_only}. When more sites are taken, this identification is no longer possible. Namely, the trace distance between states $\tilde{\rho}_3(j,j+1,j+2)$ in the bulk and the closest thermal state $\rho_3(j,j+1,j+2)$ is $\approx4\times10^{-4}$; in addition, for states of four sites, the corresponding trace distance is $\approx1\times10^{-3}$. Indeed, several expectation values of the $\tilde{\rho}_n$ states with $n>2$ are not well reproduced by thermal states $\rho_n$, and thus there is no thermalization. For nonintegrable systems this is not the case. We have verified, for $B=0.4$ and $\Delta=1$, that the expectation values of the reduced density operators of three and four sites are still well reproduced by thermal states, with corresponding trace distances across the system of $O(10^{-5})$. Thus  the conclusion of local thermalization for nonintegrable cases is robust to considering more than two sites.

\subsection{Scaling with system size}

\begin{figure}[t]
\begin{center}
\includegraphics[scale=0.95]{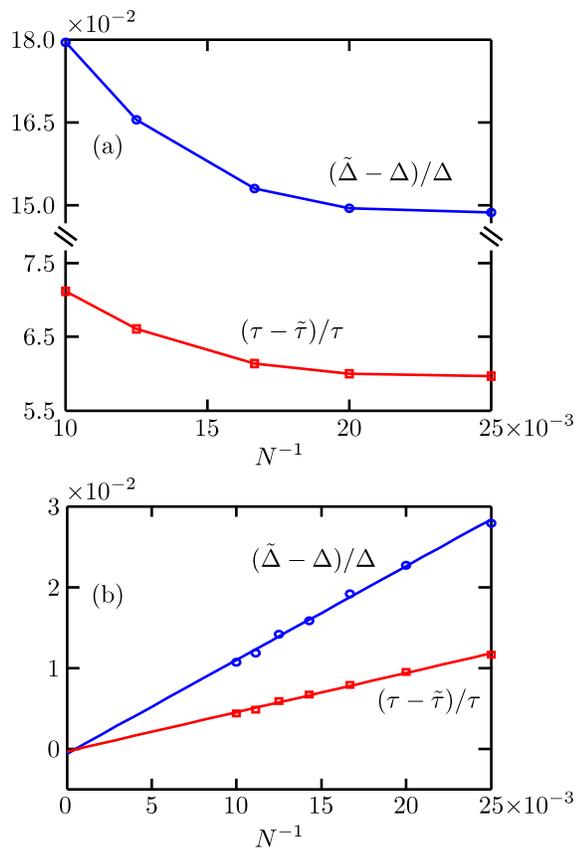}
\caption{\label{eff_couplings_scaling} (Color online) Scaling of the difference of effective and real exchange coupling ($\scriptstyle\square$) and anisotropy ($\circ$) with the system size, for the central sites and $\Delta=1.5$. (a) For integrable regime, $B=0$. The solid lines are guides to the eye. (b) For nonintegrable regime with $B=0.15$. The solid lines are the corresponding linear fits: $(\tilde{\Delta}-\Delta)/\Delta=1.16(8)N^{-1}-6(13)\times10^{-4}$, and $(\tau-\tilde{\tau})/\tau=0.49(4)N^{-1}-3(6)\times10^{-4}$. The calculations correspond to the parameters of Fig.~\ref{temp_nonequil_bvar}.}
\end{center}
\end{figure}

Finally, we discuss the effect of the system size on its local description by means of thermal states. For various sizes $N$, we obtained the effective parameters $\tilde{\tau}_j$ and $\tilde{\Delta}_j$ of the local Hamiltonian of Eq.~\eqref{local_hami_effective} for the central pair of spins, and calculated their difference to the actual Hamiltonian parameters. We consider first the integrable regime $B=0$. As shown in Fig.~\ref{eff_couplings_scaling}(a), the effective couplings diverge from the couplings of the parent Hamiltonian as $N$ increases. This provides further evidence that in the integrable limit, the system does not locally thermalize for any size, since as it becomes larger, a local description by a thermal state of the underlying Hamiltonian becomes increasingly worse.

The results are entirely different for the nonintegrable regime, even for a weak staggered magnetic field. This is illustrated for the two central spins and $B=0.15$ in Fig.~\ref{eff_couplings_scaling}(b). Notably, as the size of the system increases, $\tilde{\tau}$ and $\tilde{\Delta}$ approach $\tau$ and $\Delta$ respectively. For the sizes attainable with our numerical simulations, this approach is very well approximated by $N^{-1}$, as indicated by the fits shown in Fig.~\ref{eff_couplings_scaling}(b). This is consistent with having $\tilde{\tau}=\tau$ and $\tilde{\Delta}=\Delta$ in the thermodynamic limit, since the errors of the size-independent term of the fits are larger than their actual value, as indicated in the caption of the figure. These scaling results show that systems close to integrability will tend to a local thermal description given by their underlying Hamiltonian for sufficiently large sizes.

For the regime of parameters considered, we have established coincident transport and thermalization phenomena, depending on the integrability of the model. Namely, ballistic energy transport occurs in the integrable regime, where the system displays a total absence of local thermalization, while diffusive energy transport and local thermalization emerge in the nonintegrable regime. For the system sizes simulated, the former is clearly identified for weak staggered magnetic fields while the latter is not. However, a scaling analysis of the local properties of the NESS suggests that even there, energy diffusion and local thermalization occur simultaneously for very large systems. Whether the transition between the two transport and local thermalization regimes occurs arbitrarily close to integrability remains an open question.

\section{Conclusions} \label{conclu}

In the present work we studied the NESS of high-temperature thermally-driven one-dimensional spin-$1/2$ $XXZ$ chains, obtained by efficient matrix product simulations. We focused on two distinct phenomena, namely energy diffusion and local thermalization, which simultaneously arise from the integrability breaking of the Hamiltonian.

Specifically, we first analyzed the energy transport supported by the system when different temperatures are imposed at its boundaries by means of a two-site driving. The results show that the integrable $XXZ$ model features ballistic energy transport. On the other hand, when the integrability of the Hamiltonian is broken by means of a staggered magnetic field, the energy transport becomes diffusive. Our results thus provide new evidence to support this picture of energy transport.

Subsequently we studied the emergence of local thermal states in the same thermally-driven systems.
We observed that deep in the nonintegrable regime the system is locally described by thermal states of the underlying Hamiltonian. Close to integrability this local description does not hold for the system sizes attainable with our simulations. However, a scaling analysis with $N$ suggests the emergence of local thermalization for very large sizes in such a regime. Finally, in the integrable limit the system is not well described by local thermal states of the underlying Hamiltonian (except for a few symmetric limits). In fact, this description becomes worse as the system size increases.

These results represent the first concrete connection between the integrability of a Hamiltonian and the emergence of corresponding local thermal states in a global nonequilibrium setup. More importantly, they suggest a close connection between transport and thermalization properties. This has been recently established for integrable closed systems~\cite{mierzejewski2014prl}. Here we show, for open boundary-driven configurations, that energy diffusion and local thermalization emerge in the same (nonintegrable) regime for large chains, the latter being more susceptible to the system size. Thus it is natural to expect that an intimate relation between the two phenomena exists. A rigorous proof of such a relation is still required.

We conclude by commenting on a connection between our results and a recent numerical study of energy transport in the $XXZ$ model~\cite{karrasch2013prb}. There, two semi-infinite spin chains, initially in thermal states of different temperatures $T^{\text{L}}$ and $T^{\text{R}}$, are coupled through a single site. As the system evolves in time, the energy current at the interface between both chains saturates rapidly in the integrable limit, while it does not (in the accessible timescales) when the Hamiltonian contains staggered magnetic fields. This led to the conjecture that the relaxation of the energy current to a steady-state value would only occur for nonzero Drude weights. Additionally, even if the current at the interface reaches a steady-state value, the energy profile does not, given that the system is closed. Thus in the ballistic regime the current ``is not determined by local temperature gradients", but has the form $f(T^{\text{L}})-f(T^{\text{R}})$ for some function $f$. Our research is consistent with this observation, by indicating that in the ballistic regime it is not actually possible to provide a sensible definition of local temperatures. It would be interesting to study local thermalization in the setup of Ref.~\cite{karrasch2013prb}, or other driving schemes, to check the generality of the qualitative results we have found.

\begin{acknowledgments}
The research leading to these results has received funding from the European Research Council under the European Union's Seventh Framework Programme (FP7/2007-2013)/ERC Grant Agreement no. 319286 Q-MAC. This work was supported by EPSRC projects EP/K038311/1 and EP/J010529/1. We acknowledge Sarah Al-Assam and the TNT Library Development Team for providing the codes for the simulations carried out during our work. We gratefully acknowledge financial support from the Oxford Martin School Programme on Bio-Inspired Quantum Technologies. We thank the National Research Foundation and the Ministry of Education of Singapore for support. J. J. M.-A. acknowledges Departamento Administrativo de Ciencia, Tecnolog\'{i}a e Innovaci\'{o}n Colciencias for economic support, and L. Quiroga and F. Rodriguez for their hospitality. Finally we acknowledge F. Heidrich-Meisner and J. Nunn for valuable discussions and suggestions.
\end{acknowledgments}

\appendix

\section{Two-site thermal driving} \label{two_site_driving_app}

To drive the system out of equilibrium by a temperature imbalance, we use the so-called two-site bath operators~\cite{znidaric2010pre,znidaric2011jstat,prosen2009matrix}. These operators are designed to induce a Gibbs state of a given target temperature and chemical potential on a pair of isolated spins with Hamiltonian $h=\varepsilon_{1,2}$ and total magnetization operator $M=\sigma_1^z+\sigma_2^z$. 
So we wish to find a superoperator $\mathcal{L}_{\text{B}}(\rho)$ which satisfies the equation
\begin{equation}
\mathcal{L}_{\text{B}}(\rho_{\text{B}}) = 0,
\end{equation}
with a Gibbs state $\rho_{\text{B}}$ at temperature $T$ and chemical potential $\mu$
being the only eigenvector of $\mathcal{L}_{\text{B}}$ with zero eigenvalue, all the other eigenvalues being $-1$. This particular choice of the driving leads to the fastest convergence to $\rho_{\text{B}}$ \cite{prosen2009matrix}. To build the superoperator $\mathcal{L}_{\text{B}}$, we start by diagonalizing the thermal state of the target temperature,
\begin{equation} \label{diag_rho}
\rho_{\text{B}} = V^{\dagger}dV,
\end{equation}
with $d=\text{diag}(d_0,d_1,d_2,d_3)$ a diagonal matrix. Now we build the ``diagonal" superoperator $\mathcal{L}_{\text{B}}^{\text{diag}}$, whose only zero-eigenvalue eigenstate is $d$, i.e. 
\begin{equation} \label{eq_for_d}
\mathcal{L}_{\text{B}}^{\text{diag}}(d) = 0.
\end{equation}
If we express the matrix $d$ in the form 
\begin{equation} \label{expansion_d}
d=\frac{1}{4}\sum_{n_1,n_2=0}^{3}c_{n_1,n_2}\sigma_1^{n_1}\otimes\sigma_2^{n_2}\equiv\sum_{n=0}^{15}C_n\Omega^{n},
\end{equation}
with $\sigma^0=\mathcal{I}$ (single-site identity), $\sigma^1=\sigma^z$, $\sigma^2=\sigma^x$ and $\sigma^3=\sigma^y$, and with $\Omega^n=1/4(\sigma_1^{n_1}\otimes\sigma_2^{n_2})$ the basis elements for two sites satisfying $4\text{tr}(\Omega^{n\dagger}\Omega^{m})=\delta_{nm}$, it is easily shown that
\begin{equation}
d=C_0\Omega^0+C_1\Omega^1+C_4\Omega^4+C_5\Omega^5,
\end{equation}
with coefficients
\begin{equation}
\begin{split}
C_0&=d_0+d_1+d_2+d_3,\qquad C_1=d_0-d_1+d_2-d_3,\\
C_4&=d_0+d_1-d_2-d_3,\qquad C_5=d_0-d_1-d_2+d_3.
\end{split}
\end{equation}
Then it is easily shown that Eq. \eqref{eq_for_d} holds for the chosen basis, and the conditions specified above are satisfied, if the non-zero elements of the matrix representation of the ``diagonal" superoperator are
\begin{equation}
\begin{split}
(\mathcal{L}_{\text{B}}^{\text{diag}})_{m,m}=-1,\quad m=1,\ldots,15\\
(\mathcal{L}_{\text{B}}^{\text{diag}})_{j,0}=C_j/C_0,\quad j=1,4,5.
\end{split}
\end{equation}
We now use the ``diagonal" superoperator to define the matrix form of the superoperator $\mathcal{L}_{\text{B}}$ inducing the thermal state $\rho_{\text{B}}$. First we express $\rho_{\text{B}}$ in the $\Omega^n$ basis,
\begin{equation} \label{expansion_rho}
\rho_{\text{B}}=\sum_{n=0}^{15}\rho_n\Omega^n,
\end{equation}
with $\rho_n$ the components on each basis element. So the matrix representations of the superoperators satisfy
\begin{equation} \label{eigen_equations}
\sum_{m,n}C_m(\mathcal{L}_{\text{B}}^{\text{diag}})_{m,n}C_n=0,\quad\sum_{m,n}\rho_m(\mathcal{L}_{\text{B}})_{m,n}\rho_n=0,
\end{equation}
for $m,n=0,\ldots,15$. Using Eqs. \eqref{diag_rho}, \eqref{expansion_d} and \eqref{expansion_rho} it is shown that
\begin{equation}
C_m=4\sum_n\rho_n\text{tr}(V\Omega^nV^{\dagger}\Omega^m).
\end{equation}
Replacing this result in the left equality of Eq. \eqref{eigen_equations}, it is obtained that
\begin{equation} \label{intermediate}
\sum_{i,j}\rho_i\biggl(\sum_{m,n}\bigl(R^{\dagger}\bigr)_{im}\bigl(\mathcal{L}_{\text{B}}^{\text{diag}}\bigr)_{m,n}R_{n,j}\biggr)\rho_j=0,
\end{equation}
where we have defined the matrix elements
\begin{equation}
R_{i,j}=\frac{1}{4}\text{tr}\bigl(V^{\dagger}\Omega^iV\Omega^j\bigr).
\end{equation}
Comparing the second equality of Eq. \eqref{eigen_equations} and Eq. \eqref{intermediate}, we finally obtain
\begin{equation}
\mathcal{L}_{\text{B}}=R^{\dagger}\mathcal{L}_{\text{B}}^{\text{diag}}R,
\end{equation}
which relates the matrix forms of the ``diagonal" and complete superoperators. 

\section{Magnetothermal effects} \label{magnetothermal_section}

\begin{figure}[t]
\begin{center}
\includegraphics[scale=0.35]{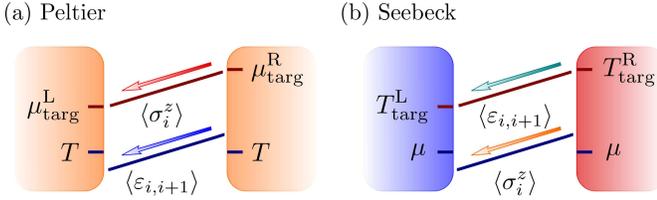}
\caption[Scheme of the magnetothermal effects induced in boundary-driven systems.]{\label{magnetothermal} (Color online) Scheme of the magnetothermal effects induced in boundary-driven systems with a finite average magnetization. (a) Peltier effect, in which an energy current is induced by a magnetic imbalance ($\mu_{\text{targ}}^{\text{R}}>\mu_{\text{targ}}^{\text{L}}$) with no temperature imbalance ($T_{\text{targ}}^{\text{L}}=T_{\text{targ}}^{\text{R}}=T$). (b) Seebeck effect, in which a spin current is induced by a temperature imbalance ($T_{\text{targ}}^{\text{R}}>T_{\text{targ}}^{\text{L}}$) with no magnetization imbalance ($\mu_{\text{targ}}^{\text{L}}=\mu_{\text{targ}}^{\text{R}}=\mu$). The arrows indicate spin and energy currents. The solid lines represent magnetization and energy profiles.}
\end{center}
\end{figure}

Here we briefly show the range of physics accessible with the two-site driving scheme. In particular we examine the emergence of magnetothermal effects, depicted in Fig.~\ref{magnetothermal}, motivated by the recent implementation of a thermoelectric heat engine in a boundary-driven configuration of ultracold atoms~\cite{brantut2013sci}. We consider only the integrable limit of the Hamiltonian ($B=0$), and illustrate how the nature of these effects may depend on the form in which they are induced.

\begin{figure}[t]
\begin{center}
\includegraphics[scale=0.78]{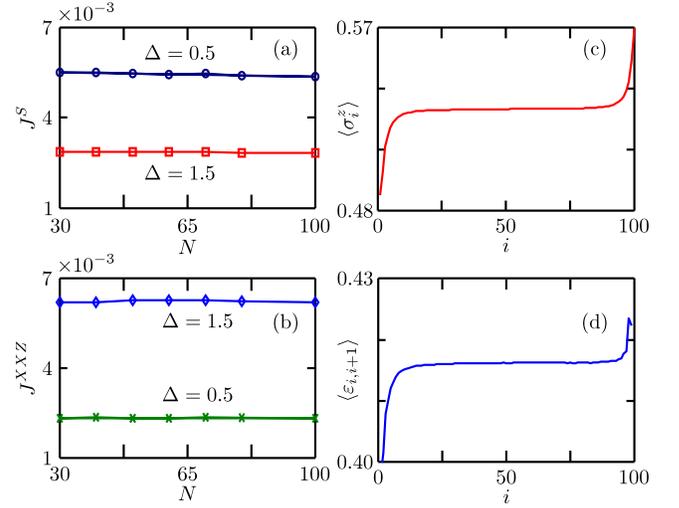}
\caption[Transport properties induced by a magnetization imbalance across an $XXZ$ chain, with finite magnetization.]{\label{peltier} (Color online) Transport properties induced by a magnetization imbalance across an $XXZ$ chain, with finite magnetization. The results correspond to $\mu_{\text{targ}}^{\text{L}}/T_{\text{targ}}^{\text{L}}=0.5$, $\mu_{\text{targ}}^{\text{R}}/T_{\text{targ}}^{\text{R}}=0.7$ and $T_{\text{targ}}^{\text{L,R}}=100$. The left panels show the spin (a) and energy currents (b), as a function of $N$. The right panels correspond to the magnetization (c) and energy profiles (d), for $\Delta=1.5$ and $N=100$.}
\end{center}
\end{figure}

We first describe how an energy current can be induced through the system in the absence of a temperature imbalance, i.e. when $T_{\text{targ}}^{\text{L}}=T_{\text{targ}}^{\text{R}}$, which corresponds to the Peltier effect. This response emerges when imposing a finite magnetization on the spin chain, which breaks the symmetry between up and down spins, in addition to a magnetization imbalance. For example, if the chemical potentials of the two boundary reservoirs satisfy $\mu_{\text{targ}}^{\text{L}}\neq\mu_{\text{targ}}^{\text{R}}>0$, a positive and homogeneous magnetization is induced in the bulk of the chain, favoring the energy current carried by spins up. A net flow of energy results, in addition to the spin current directly induced by the magnetization imbalance. As shown in Figs.~\ref{peltier}(a) and (b), these currents are independent of the size of the system, in both the weakly- and the strongly-interacting regimes. Additionally, the magnetization and energy profiles are flat in the bulk, as shown in Figs.~\ref{peltier}(c) and (d) respectively. Thus the induced spin transport is ballistic, as expected from the finite overlap between the spin and (conserved) energy current operators~\cite{zotos1997prb}, as well as the (magnetothermal) energy transport.

Importantly, a Peltier response can be induced in alternative ways. Namely, the symmetry between up and down spins could be broken by applying a homogeneous magnetic field along the system. This would induce a component of the heat current given by the product of the magnetic field and the spin current $J^{\text{S}}$, being ballistic for weak interactions ($|\Delta|<1$) and diffusive in the strongly-interacting regime ($|\Delta|>1$)~\cite{we2}. Thus the nature of the magnetothermal response of the system depends on the particular form in which it is induced.

Using the two-site driving to impose a finite and homogeneous magnetization on the system, it is also possible to induce a spin current by means of a temperature imbalance, a phenomenon known as Seebeck effect. We have verified that when $T_{\text{targ}}^{L}>T_{\text{targ}}^{R}$ and $\mu_{\text{targ}}^{\text{L}}=\mu_{\text{targ}}^{\text{R}}>0$, so there is temperature but no magnetization imbalance, the induced transport of spin and energy is ballistic for the integrable Hamiltonian, for both weak and strong interactions. Since the results have the same form than those shown in Fig.~\ref{peltier} for the Peltier effect, i.e. flat magnetization and energy profiles in the bulk and size-independent currents, they are not shown.  

These results demonstrate that under the two-site driving scheme used here,  ballistic magnetothermal responses exist in the integrable $XXZ$ model, for both weakly- and strongly-interacting regimes~\cite{louis2003prb,ajisaka2012prb}.

\section{Obtaining local temperatures and chemical potentials} \label{self_consistent}

\begin{figure}[t]
\begin{center}
\includegraphics[scale=0.95]{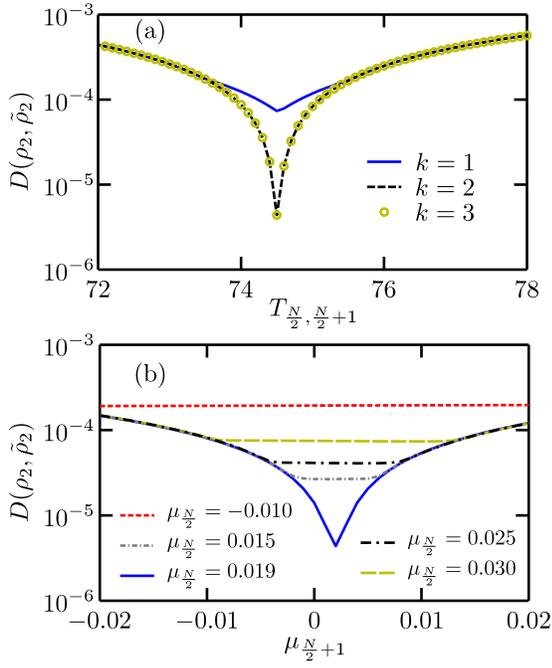}
\caption{\label{self_figure} (Color online) Trace distance between the two-site reduced density operator $\tilde{\rho}_2(\frac{N}{2},\frac{N}{2}+1)$ and two-site thermal states $\rho_2$ with trial temperatures $T_{\frac{N}{2},\frac{N}{2}+1}$ and chemical potentials $\mu_{\frac{N}{2}}$ and $\mu_{\frac{N}{2}+1}$. The results correspond to a system with $B=0.4$ and the parameters of Fig.~\ref{temp_nonequil_bvar}. (a) Sweep over temperature for three iterations $k=1,2,3$ of the self-consistent process. Iteration $k=1$ corresponds to the mean-field chemical potentials of Eq.~\eqref{mf_chem_pots}. (b) Sweep over chemical potential of site $\frac{N}{2}+1$ for various fixed potentials of site $\frac{N}{2}$, for iteration $k=1$. No more iterations are depicted here, since they give the same trace distances than those of $k=1$.}
\end{center}
\end{figure}

In Section~\ref{determination_thermal} we briefly described how to determine local temperatures and chemical potentials of the thermally-driven spin chain. Now we present more details of this self-consistent calculation. 

We start by comparing the two-site reduced density operator $\tilde{\rho}_2(j,j+1)$ of each pair of neighboring sites with thermal states $\rho_2$ of the form in Eq.~\eqref{grand_can_state_local}, with the mean-field chemical potentials of Eq.~\eqref{mf_chem_pots} and trial temperatures within a range $[T_{\text{min}},T_{\text{max}}]$, separated by a step $\delta T$. Then we select the temperatures $T_{j,j+1}$ that minimize the trace distance of Eq.~\eqref{trace_dist} for each pair. This step is exemplified for the central sites of a particular spin chain in Fig.~\ref{self_figure}(a) (blue solid line); in this case, the temperature selected is $T_{\frac{N}{2},\frac{N}{2}+1}=74.51$. Subsequently, we compare states $\tilde{\rho}_2(j,j+1)$ to thermal states $\rho_2$ with the selected temperatures and trial chemical potentials $\mu_j$ and $\mu_{j+1}$ within a range $[\mu_{\text{min}},\mu_{\text{max}}]$ (separated by a step $\delta\mu$), again by means of the trace distance. This is illustrated in Fig.~\ref{self_figure}(b) for the central sites of the chain, where for different fixed values of $\mu_{\frac{N}{2}}$ we show the corresponding trace distances for a sweep over $\mu_{\frac{N}{2}+1}$. We then select the values $(\mu_{\frac{N}{2}},\mu_{\frac{N}{2}+1})$ that minimize $D(\rho_2,\tilde{\rho}_2)$; in the example they are $(0.019,0.002)$. This corresponds to the first iteration ($k=1$) of the process. Afterwards, with the obtained values of chemical potentials, we select a new temperature for each pair of sites by means of the same process, and then we identify new values of chemical potentials. This corresponds to the second iteration ($k=2$), for which a large decrease of the minimal trace distances is observed with respect to the first iteration (see black dashed line of Fig.~\ref{self_figure}(a)). The procedure is repeated until the values of temperatures and chemical potentials remain unaltered when increasing the number of iterations, up to the accuracy given by the steps $\delta T$ and $\delta\mu$ selected. In the example of Fig.~\ref{self_figure} this has been already achieved with the third iteration ($k=3$), for which the values of $D(\rho_2,\tilde{\rho}_2)$ are the same than those of the second iteration (see the symbols of Fig.~\ref{self_figure}(a)).

\bibliography{mybib_heat}

\end{document}